\documentclass[doublecol]{epl2}

\usepackage{amsmath,amssymb}
\usepackage{graphicx}
\usepackage{dcolumn}
\usepackage{bm}

\let\a=\alpha \let\b=\beta    \let\g=\gamma     \let\d=\delta     \let\e=\varepsilon
  \let\h=\eta     \let\th=\vartheta \let\k=\kappa     \let\l=\lambda
\let\m=\mu    \let\n=\nu              \let\p=\pi        \let\r=\rho
\let\s=\sigma \let\t=\tau            
   \let\o=\omega     
 \let\D=\Delta

\let\io=\infty

\let\e=r

\def\be{\begin{equation}}
\def\ee{\end{equation}}
\def\bea{\begin{eqnarray}}\def\eea{\end{eqnarray}}
\def\bean{\begin{eqnarray*}}\def\eean{\end{eqnarray*}}

\def\nn{\nonumber}

\def\VV{{\cal V}}

\def\RR{{\cal R}}\def\LL{{\cal L}} 
\def\j{\jmath}

 \def\pp{{\bf p}}
\def\xx{{\bf x}} \def\yy{{\bf y}} 
\def\kk{{\bf k}}

\usepackage[makeroom]{cancel}
\usepackage{color}

\title{Non-integrable fermionic chains near criticality.}
\author{F. Bonetto\inst{1} and V. Mastropietro\inst{2}}
\institute{
\inst{1} School of Mathematics, Georgia Institute of Technology, Atlanta, 
GA 30332, USA.\\
\inst{2} Dipartimento di Matematica, Universit\`a di Milano, Via Saldini 50, 
Milano, Italy}

\abstract{
We compute the Drude weight and the critical exponents as functions of the 
density in non-integrable generalizations of XXZ or Hubbard chains, in the 
critical zero temperature regime where Luttinger liquid description breaks down 
and Bethe ansatz cannot be used. Even in the regions where irrelevant terms 
dominate, no difference between integrable and non integrable models appear in 
exponents and conductivity. Our results are based on a fully rigorous two-regime 
multiscale analysis and a recently introduced partially solvable 
model.}


\begin{document}
\maketitle
\section{Introduction} 
Understanding whether the behavior of exactly solvable models is generic and 
persists in presence of integrability-breaking terms is a central issue in 
physics. Interacting fermionic chains provide an ideal arena, thanks to the 
presence of Bethe ansatz solvable models, like the XXZ or the Hubbard model, and 
the fact that cold atoms allow, at least in principle, an experimental 
verification \cite{1,2,2xx}. Exact solutions provide a rather complete picture, 
including critical exponents at zero temperature for all densities \cite{ko}, 
Mazur bounds for Drude weights (whose finiteness signals an infinite 
conductivity) at finite temperature \cite{5,6} and dynamical correlations 
\cite{6aa,gla,gla1}. In addition, Drude weights can be obtained via dynamical 
evolution of partitioned systems  \cite{1a,2a,3a,4a,5a,6a,7a,8a,9a}. 

Luttinger liquid theory \cite{Ha} predicts the behavior of the Luttinger 
model to be generic for non-integrable systems \cite{M1}. This was rigorously 
proved \cite{BFM} for static zero temperature properties around the half filled 
band case, where the dispersion relation is essentially linear. These 
limitations are necessary; solvable models show that non linear dispersion 
relations produce behaviors different from that of the Luttinger model in the 
dynamical correlations or at finite temperature; the same is true for static 
zero temperature properties at low or high densities. 

For the same non linear 
lattice dispersion relation, integrable or non integrable interactions differ by 
irrelevant terms, usually neglected in field theoretic Renormalization Group 
(RG) analysis; for instance, the addition of a next to nearest neighbor 
interaction makes the XXZ model not solvable. The RG irrelevance of these terms 
does not make them unimportant. On the contrary, it has been proposed that at 
positive temperature the Drude weight can depend dramatically on the 
integrability of the interaction \cite{5,6}, in analogy with the classical case 
\cite{BoL}. This scenario still lacks confirmation 
\cite{8,10,12,13,13a,15}. More generally, irrelevant terms are known
to play a crucial role for transport properties. For instance, they ensure the 
cancellation of all the interaction corrections of the optical conductivity of 
graphene\cite{x1,x2}.

The natural question we address here is the following: for which properties is 
the behavior found in Bethe ansatz solvable models generic even when the 
Luttinger description breaks down and  physics is dominated by irrelevant terms? 
We answer this question in the case of static zero temperature properties in the 
low or high density regions, away from Luttinger linear behavior. This is 
achieved via a two-regime non-perturbative RG scheme that keeps fully into 
account irrelevant terms. In the second regime, in the spinful case we exploit 
emerging symmetries by using a recently introduced QFT model \cite{BFM2} with a 
RG flow exponentially close to the flow of the non integrable chains. This QFT 
model is partially solvable in the sense that only the density correlations can 
be obtained in closed form. 

We find that the critical exponents, in the low or high density limit, tend to 
their non interacting value in the spinless case, while in the spinful case 
their limiting value depends strongly on the interaction. In both cases the 
Drude weight behave essentially as in the non interacting case. In the special 
case of solvable interactions, Bethe ansatz results are recovered. 

Our analysis 
shows that there is no qualitative difference between integrable and non 
integrable models at zero temperature; the exponents have a similar behavior and 
the conductivity is infinite, and the limiting value of the Drude weight is the 
same. This is true notwithstanding the fact that near criticality the physics 
is completely dominated by the irrelevant terms. As soon as $T\not=0$, it is 
common belief that a completely different 
behavior of the Drude weight occurs, being it vanishing or not depending on 
integrability.
The fact that integrability breaking perturbations are unable to produce any 
difference at $T=0$,
even in regions where irrelevant terms dominate,
makes it also possible a scenario in which a difference is visible 
only at not too small temperatures.

\section{Main results}
We consider a model of interacting fermions with Hamiltonian
\begin{align}
H=&-\frac12\sum_{x,\s} 
(a^+_{x,\s} a^-_{x+1,\s}+c.c.)-\m \sum_{x,\s} a^+_{x,\s} a^-_{x,\s}\nn\\
&+\l\sum_{\substack{ x, y \\  \s,\s'}}  w(x-y) 
a^+_{x,\s}a^-_{x,\s} a^+_{y,\s'}a^-_{y,\s'}\label{1}
\end{align}
where $a^\pm_{x,\s}$ are fermionic creation or annihilation operators, $\s$ is 
the spin ($\s=0$ in the spinless case and $\s=\uparrow,\downarrow$ in the 
spinning case), $x$ are points on a one dimensional lattice and $w(x)$ is a 
short range potential such that $\sum_x |x|^\a |w(x)|<\io$ for some $\a>0$. In 
the spinless case with $w(x-y)=\d_{x,y+1}$ the system reduces to the $XXZ$ model 
and in the spinning case with $\l w(x-y)=U\d_{x,y}$ it reduces to the Hubbard 
model. For other choices of the interaction no solution is known. 

The truncated Euclidean correlations are 
\begin{equation*}
\langle 
O_{\xx_1}...O_{\xx_n}\rangle=\langle 
\mathbf{T}(O_{\xx_1}...O_{\xx_n})\rangle_{T}\, ,
\end{equation*}
where $\mathbf{T}$ is the time ordering operator, $\xx=(x_0,x)$, $O_\xx=e^{H 
x_0}O_x e^{-H x_0}$ and $\langle 
\cdot \rangle_T$ are the thermodynamic truncated averages. Finally 
$S(\xx-\yy)=\langle a^-_\xx a^+_\yy\rangle$ denotes the 2-point correlation 
function. 

The density is $\r_x=\sum_\s a^+_{x,\s}a^-_{x,\s}$ and the current is 
defined via the continuity equation that gives 
\begin{equation*}
j_x=\frac{1}{ 
2i}\sum_\s(a^+_{x+1,\s}a^-_{x,\s}+a^+_{x,\s}a^-_{x+1,\s})\,.
\end{equation*}
Writing $\pp=(p_0,p)$, the (Euclidean) zero temperature Drude weight $D$ and 
the susceptibility $\k$ are given by  
\begin{equation*}
\k=\lim_{p\to 0} \lim_{p_0\to 0} \langle\hat 
\r_\pp \hat \r_{-\pp}\rangle_T\quad\mathrm{and}\quad
D=\lim_{p_0\to 0} \lim_{p\to 0} D(\pp)
\end{equation*}
with $D(\pp)=\langle\hat j_\pp \hat j_{-\pp}\rangle_{T}+\D$ and 
\begin{equation*}
\D=-\frac{1}{2} \sum_\s \langle a^+_{x,\s} a^-_{x+1,\s} 
+a^+_{x+1,\s}a^-_{x,\s}\rangle\,.
\end{equation*}
Here $\hat f(\pp)$ represents the Fourier 
transform of $f(\xx)$. A Ward Identity (WI) gives $p_0^2\langle\hat \r_\pp \hat 
\r_{-\pp}\rangle =4\sin^2 p/2 D(\pp)$ which implies that 
\begin{equation*}
\lim_{p\to 0} 
\lim_{p_0\to 0} D(\pp)=\lim_{p_0\to 0}\lim_{p\to 0}\langle\hat \r_\pp \hat 
\r_{-\pp}\rangle=0\,. 
\end{equation*}
Note that $D(\pp)$ is not continuous at $\pp=0$ and it is 
essential to take the limits in the correct order. Moreover the limit $p_0\to 0$ 
should be taken along the imaginary axis, but Wick rotation holds for this model 
\cite{MP}. 

\smallskip

\noindent{\bf Theorem}\ {\it Consider the Hamiltonian \eqref{1} with 
$\mu=\mu_R+\nu(\lambda,r)$ and $\mu_R=-\cos p_F= \pm 1\mp \e$. 
Then we have 
\begin{equation*}
D=\frac{K v}{\pi}\quad\mathrm{and}\quad\k= \frac{K}{\pi v}
\end{equation*}
where:
\smallskip

\noindent$\bullet$ In the spinless case for $|\l|$ small we have 
\begin{equation*}
\nu(\lambda,r)=2\l \hat w(0) \frac{p_F}{\pi}
+O(\l\e)
\end{equation*}
while 
\begin{align*}
K=&\frac{1-\t}{1+\t}\,,\\
v=&\sin p_F(1+O(\l\e^\th))\,,\\
\t=&\l\frac{\hat w(0)-\hat w(2p_F)}{2\pi v} +O(\l^2\e^{\th}) \, ,
\end{align*}
with $\th\in (1/3,1/2)$;
\smallskip

\noindent$\bullet$ In the spinful case for $\tilde\l = 
\frac{\l}{\sin p_F}\geq 0$ small we have 
$\nu(\lambda,r)=O(\tilde\l\sqrt{\e})$ while
\begin{equation*}
K=\sqrt{\frac{(1-2\n_\r)^2-\n_4^2}{ (1+2\n_\r)^2-\n_4^2  }}\quad v^2=\bar v^2 
\frac{(1+\n_4)^2-4\n_\r^2}{ (1-\n_4)^2-4\n_\r^2}
\end{equation*}
where 
\begin{align*}
\bar v=&\sin p_F(1+O(\tilde\l\e^\th)+O(\tilde\l^2))\,,\\ 
\n_4=&\tilde\l \frac{\hat w(0)}{2\pi} +O(\tilde\l^2)\,,\\ 
\n_\r=&\frac{\tilde\l}{2\p }(\hat w(0)-\frac{\hat w(2p_F)}{2})+O(\tilde\l^2)\, .
\end{align*}

\smallskip

\noindent In both cases, $S(\xx-\yy)$ decays for large distance as
$|\xx|^{1+\h}$ with $2\h=K+K^{-1}-2$.
}
\smallskip

In the Theorem $\e$ is a parameter that measures the distance of $\mu$ from the 
critical chemical potential $\mu_c$. In the spinless case $\m_c$ is shifted by 
the interaction and we get $\m_c=1+2\l \hat w(0)$ for $\mu_R=1$ and $\m_c=-1$ 
for $\mu_R=-1$. In the $XXZ$ chain $h_c+\l=\m_c$. When $\e\to 0$ we get $K\to 1$ 
and $D/\sin p_F\to \frac{1}{\pi}$, that is the critical exponent and the Drude 
weight tend to their non-interacting values. Fig. 1 shows the behavior of $D$ 
and $K$ as function of the density close to the critical point; in the XXZ case 
it closely reproduces the features found by the exact solution, see e.g. Fig. 1 
in \cite{S1} or Fig. 1 in \cite{PZ}.
\setlength{\unitlength}{1cm}

\begin{figure}
\centerline{\includegraphics[width=0.53\textwidth]{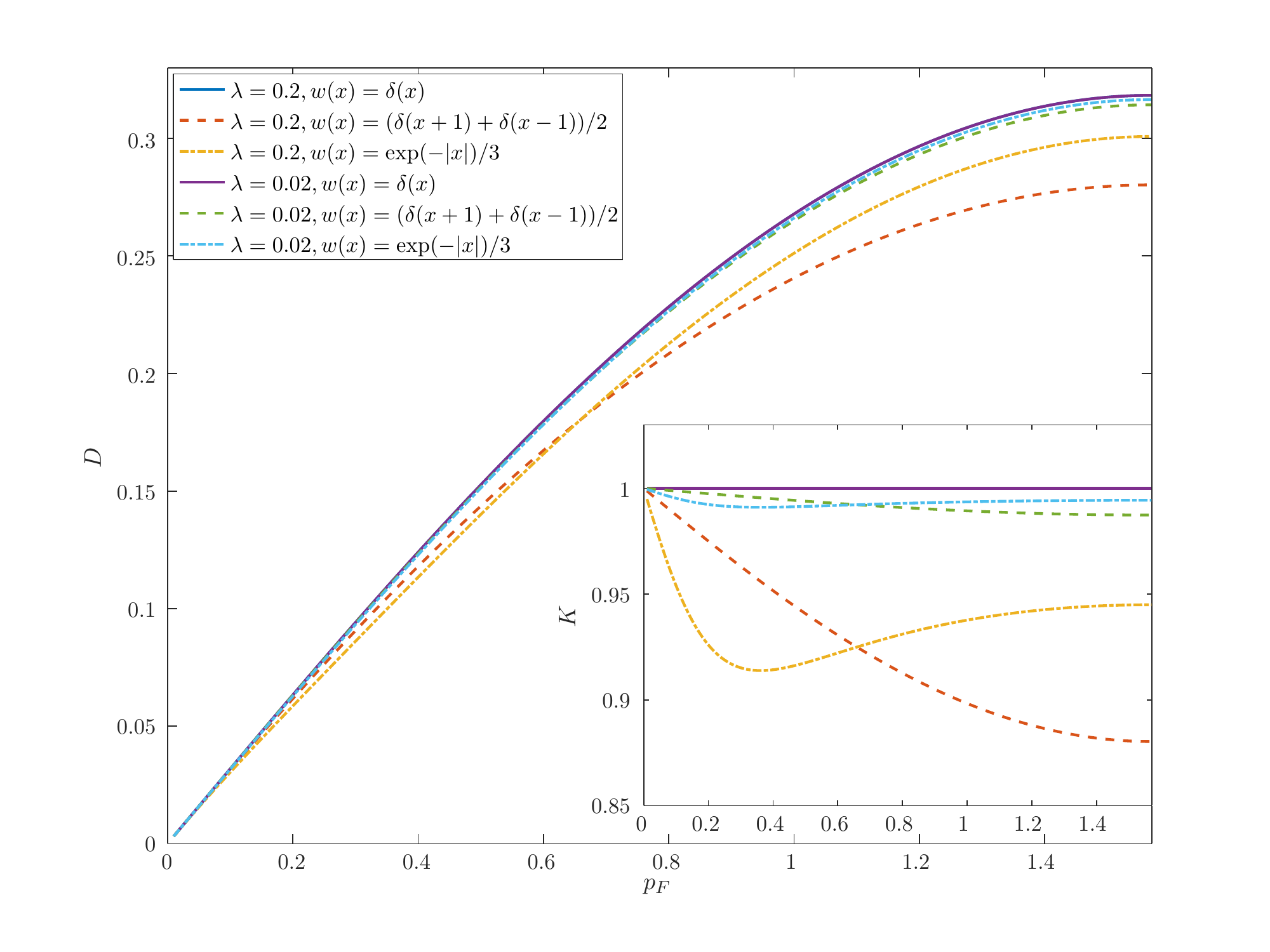}}
\caption{The main graph is the Drude weight $D$ at fixed
$\l$ in the spinless case. The inset shows $K$.}
\label{F1} 
\end{figure}

\begin{figure}
\centerline{\includegraphics[width=0.53\textwidth]{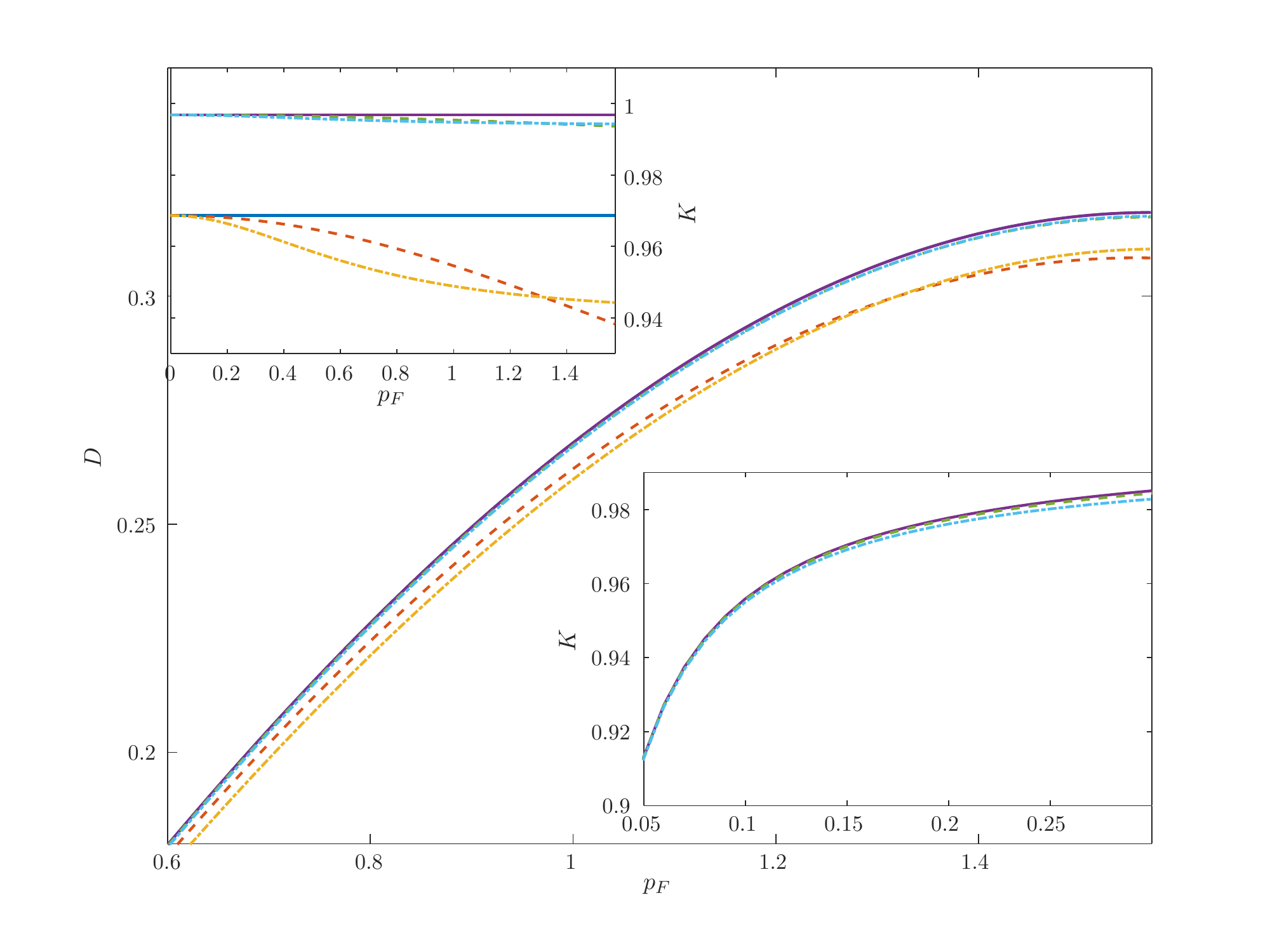}}
\caption{ The main graph is the Drude peak $D$ at fixed
$\l$ for the spinful case. 
The upper insets shows $K$ at fixed $\tilde\l$ while the lower inset 
shows $K$ as a function of $\l$. Colors and dashes are as 
in Fig. \ref{F1}}
\label{F2}
\end{figure}

In the spinful case we rescale the interaction as $\l=\tilde\l \sin p_F$. In 
term of $\tilde\l$ our results hold uniformly in $\e$. In contrast with the 
spinless case, the theory is strongly interacting since at criticality we have 
$K\to 1-\tilde\l \hat w(0)/\pi+O(\tilde\lambda^2)$. A remarkable cancellation 
takes place in the Drude weight and $D$ behaves as in the non interacting case 
when $\e\to 0$ (at least up to $O(\tilde\l^2)$ terms), that is 
\begin{equation*}
\frac{D\pi}{\bar v}=\frac{1+\n_4-2\n_\r}{1-\n_4+2\n_\r}\sim 1
\end{equation*}
for $\e\sim 0$. Such a behavior is 
present in the Hubbard model, but it is proven here to be a generic feature. It 
was missed in previous attempts based of field theoretic RG methods. Fig. 
\ref{F2} shows the behavior of $D$ and $K$ for integrable and non integrable 
interactions, as function of $\l$ and $\tilde\l$. In the Hubbard case Fig.  
\ref{F2} reproduces Bethe ansatz result (e.g. Fig. 9.2, 9.3  of \cite{ko} or 
Fig. 13, 14 of \cite{S}).

\section{RG analysis: the quadratic regime} We write the Euclidean 
correlations in terms of a Grassmann integral 
\begin{equation*}
e^{W(A,\phi)}=\int P(da)e^{-\VV-\nu N+B(A,\phi)} 
\end{equation*}
where $P(da)$ is a Grassmann integration on the 
Grassmann algebra generated by the variables $a^\pm_{\xx,\sigma}$ with 
propagator 
\begin{equation*}
g(\xx-\yy)=\frac{1}{4\pi^2}\int e^{-i\kk(\xx-\yy)}\hat 
g(\kk)\,d\kk
\end{equation*}
where
\begin{equation*}
\hat g(\kk)=\frac{1}{-ik_0-\cos k+\cos p_F}\, .
\end{equation*}
Moreover $\VV$ is the interaction and $\nu N= 
\nu\int d\xx\, a^+_{\xx,\s}a^-_{\xx,\s}$ is a counterterm introduced to take 
into account the renormalization of the chemical potential, that is we write 
$\m=\m_R+\n$ with $\m_R\equiv \cos p_F$. Finally $B(A,\phi)$ is a source term. 
Differentiating $W(A,\phi)$ with respect to $\phi$ produces correlations of 
fermionic fields, while differentiating with respect to $A$ produces 
correlations of currents or densities. 

\begin{figure}[t]
\centerline{\includegraphics[width=0.4\textwidth]{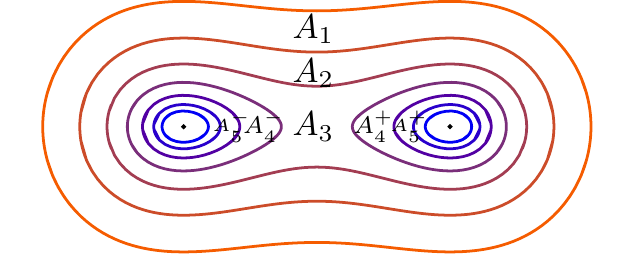}}
\caption{Schematic representation of the support $A_h$ of the propagator
$g^{(h)}(\kk)$ as a function of $h$.}
\label{F3} 
\end{figure}

The starting point of the RG analysis is 
the decomposition 
\begin{equation}\label{decomp}
\hat g(\kk)=\sum_{h=-\io}^1 \hat f_h(\kk) \hat 
g(\kk)=\sum_{h=-\io}^1 \hat g^{(h)}(\kk)
\end{equation}
where $\hat f_h(\kk)$ is a compact 
support 
function non vanishing only for $\sqrt{k_0^2+(\cos k-\cos p_F)^2}\sim 2^{h}$, 
see Fig. \ref{F3}. From eq.\eqref{decomp} and the prperties of Grassmanian 
integrations we have that we can write $a^\pm_{\xx,\sigma}=\sum_{h=-\io}^1 
a^{h,\pm}_{\xx,\sigma}$ with $P(da)=\prod_{h=-\io}^1 P(da^h)$. This 
decomposition naturally leads to identify two regions, separated by the energy  
scale $2^{h^*} \sim \e$; in the region where the energy is greater $r$ the 
dispersion relation is essentially quadratic, while for smaller energies it is 
essentially linear with a slope of $\sin p_F\sim \sqrt{\e}$. 

In the high energy region where $h\ge h^*$  the single scale propagator 
satisfies the scaling relations $g^{(h)}(\xx)\sim 2^{h/2}g^{(0)} (2^h x_0, 
2^{h/2} x)$ and the scaling dimension is $D^1=3/2-l/4-m/2$ where $l$ is the 
number of $a$ fields and $m$ the number of $A$ fields. 

We focus on the $A=\phi=0$ case. We  define the effective potential on scale $h$ 
recursively as 
\begin{equation*}
V^h(a^{\leq h})=\log \int P(da^h)e^{V^{h+1}(a^{\leq h+1})}
\end{equation*}
where $a^{\leq h}=\sum_{k=-\infty}^h a^{k}$. It can be written as 
\begin{equation*}
V^h=\LL V^h+\RR V^h\,,
\end{equation*}
where $\RR V^h$ is sum of all irrelevant terms, that is monomials 
in the fields with $D^1<0$ while 
\begin{align}
\LL& V^h=2^{\frac{h}{2}}\l_h F_\l+ 2^h\n_h\sum_\sigma\int d\xx \,a^{+,\leq 
h}_{\xx,\s} a^{-,\leq h}_{\xx,\s}+\label{ss}\\
& i_h \sum_\sigma\int d\xx \,a^{+,\leq h}_{\xx,\s}
 \partial_0 a^{-,\leq h}_{\xx,\s}+
 \d_h \sum_\sigma\int d\xx \,a^{+,\leq h}_{\xx,\s}
 \partial^2 a^{-,\leq h}_{\xx,\s}\nonumber
\end{align}
where 
\begin{equation*}
F_\l=\int d\xx \,a^{+,\leq h}_{\xx,\uparrow} a^{-,\leq 
h}_{\xx,\uparrow}a^{+,\leq h}_{\xx,\downarrow} a^{-,\leq h}_{\xx,\downarrow}
\end{equation*}
in the spinful case and $F_\l=0$ if the fermions are spinless. Notice
the absence of the term $\int a^+_{\xx,\s} \partial a^-_{\xx,\s}d\xx$ and of 
local terms with six fields due to parity and the Pauli principle, 
respectively. 

After integrating the field $a^h$ we obtain $V^{h-1}$ as a sum of monomials in 
the fields, that is
\begin{equation*}
 V^{h-1}(a^{\leq h-1})=\int W_l^{h-1}(\xx_1,\ldots,\xx_l)\prod_{i=1}^l a^{\leq 
h-1}_{\xx_i}
\end{equation*}
where 
$W^{h-1}_l$ is expressed as a series in the running coupling constant (r.c.c.) 
$\boldsymbol{\eta}_h=(\n_k, i_k,\d_k,\l_k)$ (with $\l_k\equiv 0$ in the spinless 
case), $k\ge h$. We can now write $V^{h-1}=\LL V^{h-1}+\RR V^{h-1}$ as in 
(\ref{ss}) with $h-1$ replacing $h$ and use the local terms to compute the 
r.c.c. on scale $h-1$. This produces an expansion of the kernels $W_l^h$ in in 
terms of the r.c.c.. Calling $\epsilon_h=\max_{k>h} |\boldsymbol{\eta}_k|$, we 
get 
\begin{equation*}
|| W_l^{h-1}||\le 2^{h(3/2-l/4)}\sum_{n}  C^n \epsilon_h^n
\end{equation*}
Convergence in the r.c.c. 
follows from determinant bounds \cite{BoM}, which imply convergence in $\l$ if 
the r.c.c. remain close to their initial value during RG iteration.

The above construction gives the recursive relation 
\begin{equation*}
\boldsymbol{\eta}_{h-1}=\boldsymbol{\eta}_{h}+\beta^h(\boldsymbol{\eta}_{h} , 
\ldots,\boldsymbol{\eta}_0)\,. 
\end{equation*}
The flow generated by $\beta^h$ can be analyzed 
rigorously as in \cite{BoM}. The main observation is that at 
$\e=0$ all graphs with a closed fermionic loop vanish while the tadpole graph 
gives the shift of the chemical potential. Therefore in the spinless case we get 
$|i_h|,|\d_h|\le C\e^\th |\l|$ where the factor $\e^\th$ due to the irrelevance 
of the quartic terms. Similarly the contribution to  $\n$ are the tadpole graph 
plus $O(\l \e)$.

In the spinful case we must also consider $\l_h$ which obeys the recursive 
relation $\l_{h-1}=2^{\frac 12}\l_h-a \l_h^2+O(\l_h^3)$ with $a>0$, from which 
$|\l_{h^*}|\le  C|\tilde\l|$. We thus see a non trivial fixed point that lie 
outside our convergence radius. For the other r.c.c. we get 
$|i_{h^*}|,|\d_{h^*}|\leq C\e^\th |\l|+O(\tilde\lambda^2)$ while the 
contribution to  $\n$ are the tadpole graph plus $O(\l \e)+O(\tilde\lambda^2)$, 
see also Appendix A. 
This is due to the lack of the dimensional gains of the spinless case 
for graphs of higher order.

\section{RG analysis: the linear regime} After the integration of the 
fields $a^1,a^0,\ldots,a^{h^*}$ we arrive to a functional integral of the form 
$\int P(da^{\le h^*}) e^{-\VV^{h^*}(a)}$, where $P(da^{\le h^*})$ has a 
propagator that depends only on the momenta in two disconnected regions around 
the 2 Fermi points $(0,\pm p_F)$, see Fig. \ref{F3}. Therefore we write $a^{\le 
h^*}$ as sum of 2 independent fields 
\begin{equation*}
a^{\le h^*}=\sum_{\o=\pm} e^{i\o p_F 
x}a^{\le h^*}_{\o,\xx}
\end{equation*}
with propagator 
\begin{equation*}
\hat g^{(\le h^*)}_\omega(\kk)=\frac{\tilde f_{\le h^*}(\kk)}{ -i k_0+\o 
v_{h^*} k}+\hat r^{h^*}(\kk)\,,
\end{equation*}
where $v_{h^*}=O(\sqrt\e)$, and  $\tilde f_{\le 
h^*}(\kk)$ 
is 
different from 0 only if $k_0^2+v_{h^*}^2k^2\leq 2^{h^*}$. Finally $\hat r(\kk)$ 
is a bounded correction. In this case the scaling dimension is  $D^2=2-l/2$; we 
write again $V^h=\LL V^h+\RR V^h$, where $\RR V^h$ contains all terms with 
negative scaling dimension while $\LL V^h$ contains $\n_h$, the renormalization 
of the chemical potential, and the quartic terms (quadratic marginal terms 
produce the wave function renormalization $Z_h$ and the renormalized Fermi 
velocity $v_h$). In the spinless case  the quartic local terms have the form 
$\l_h \int d\xx a^{+,\le h}_{\xx,+}  a^{-,\le h}_{\xx,+} a^{+,\le h}_{\xx,-}  
a^{-,\le h}_{\xx,-}$ with
\begin{align}
&\l_{h^*}=\l (\hat w(0)-\hat w(2 p_F))+\nn\\
&\ \ \sum_{k=h^*}^0 (W_4^{k}(p_F,p_F,-p_F,-p_F)-W_4^{k}(p_F,-p_F,-p_F,p_F))\nn
\end{align}
Due to the parity of the interaction, the first term is $O(\l \e)$ while the 
second is close to $p_F^2 \partial^2 W_4^k$.  Since 
\begin{equation*}
\sum_{k=h^*}^0 |\partial^2 
W_4^{k}| \le \sum_{k=h^*}^0 \l^2 2^{h(-1/2+\th)}\le C\l^2 \e^{-1/2+\th}
\end{equation*}
we get 
$\l_{h^*}\sim O(\l\e^{\frac12+\th})$, so that it vanishes as $\e\to 0$. In the 
spinful case there are three local quartic terms (if $p_F\not=\pi/2$): 
\begin{itemize}
\item $g_{1,h} 
\int a^+_{\xx,\o,\s}a^-_{\xx,-\o,\s}a^+_{\xx,-\o,\s'} a^-_{\xx,\o,\s'}$ with 
$g_{1,h^*}=2^{h^*/2} (2\tilde\l \hat w(2p_F)+O(\tilde\l^2))$ where the 
$2^{h^*/2}$ comes from the scaling dimension; 
\item $g_{2,h}\int 
a^+_{\xx,\o,\s}a^-_{\xx,\o,\s}a^+_{\xx,-\o,\s'}a^-_{\xx,-\o,\s'}$ with 
$g_{2,h^*}=2^{h^*/2}( 2\tilde\l w(0)+O(\tilde\l^2))$; 
\item $g_{4,h}\int 
a^+_{\xx,\o,\s}a^-_{\xx,\o,\s}a^+_{\xx,\o,\s'}a^-_{\xx,\o,\s'}$ with 
$g_{4,h^*}=2^{h^*/2}(2\tilde \l w(0)+O(\tilde\l^2))$. 
\end{itemize}
The integration over the 
time variables produces a factor $v^{-n+1}$ which is compensated by the $v^n$ of 
the coupling, so that the convergence radius (in $\l$ for the spinless case or 
$\tilde\l$ for the spinful case) is $\e$ independent. Observe that the small 
factor in the effective coupling is produced essentially by Pauli principle in 
the spinless case, while it follows from our choice $\l=\tilde\l \sin p_F$ in 
the spinful case. 

Finally we have to discuss the flow of the running coupling constants. The 
single scale propagator $\hat g^h(\kk)$ is sum of of a "relativistic" part 
\begin{equation*}
\hat {g}^h_{\o,rel} (\kk)=\frac{1}{ Z_h}\frac{\tilde f_h(\kk)}{-i k_0+\o 
v_{h} k}
\end{equation*}
and a correction $\hat 
r^h(\kk)$, smaller by a factor $ \frac{2^h}{v^2_{h^*}}$, that takes into account 
the non linear corrections to the dispersion relation. In the spinless case the 
beta functions for $\l_h$ and $v_h$ are asymptotically vanishing (i.e. the only 
contributions come from the corrections $\hat r^{h}$) while 
\begin{equation*}
|\b_\l^h|\le C 
\frac{\l^2_{h}}{v_{h^*} } \frac{2^h}{v^2_{h^*}}\quad\mathrm{ and}\quad 
|\b_\d^h|\le C \l_{h} 
\frac{2^h}{v^2_{h^*}}\,.
\end{equation*}
Thus we get $|\l_h|\le C \l \e^{1/2+\th}$ while 
$v_{-\io}=\sin p_F(1+O(\l \e^{\th}))$. Finally we have $Z_h\sim Z_{h^*} 2^{-\h 
h}$ with $\h=\h_i(\frac{\l_{-\io}}{ v_{-\io}} )$, see also Appendix B.

In the spinful case if $\l>0$, we get $g_{2,h}\to g_{2,-\io}$ and $g_{4,h}\to 
g_{4,-\io}$ with $ g_{2,-\io}=g_{2,h^*}-g_{1,h^*}/2+O(\tilde\l^2 \e^{1/2})$ and 
$g_{4,-\io}=g_{4,h^*}+O(\tilde\l^2 \e^{1/2})$. Finally we have $g_{1,h}\sim 
\frac{g_{1,h^*}}{1-a g_{1,h^*}(h-h^*)}\to 0$ as $h\to-\io$. Similarly we get 
$\bar v=\sin p_F(1+O(\tilde\l\e^{\th})+O(\tilde\l^2))$.

\section{Emerging Chiral model} Here we focus on the spinful case, since 
the spinless one is a special case of the following discussion. In the second 
regime a description of relativistic chiral fermions emerges, up to irrelevant 
terms, and one needs to exploits its symmetries. A way to do that is to 
introduce a reference model whose parameters can be fine tuned so that the 
difference between the running coupling constants of the non integrable chain 
and those of the reference model is small. 
The somewhat natural choice of the 
Luttinger model does not work, as the difference produced 
by the $g_1$ coupling vanishes in a non summable way. 

We introduce a model \cite{BFM2} of fermions $\psi^\pm_{\o,\s}$ $\o=\pm$ with 
propagator 
\begin{equation*}
\hat {g}^h_{\o,chi} (\kk)=\frac{1}{Z}\frac{\tilde f_{\leq N}(\kk)}{ -i k_0+\o v 
k}
\end{equation*}
and interaction given by 
$\VV=\bar g_{1} F_1+\bar g_{2} F_2+\bar g_{4} F_4$ where
\begin{align*}
F_1&=\frac{1}{2}\sum_{\o,\s,\s'} \int 
\tilde 
w(\xx-\yy)\psi^+_{\xx,\o,\s}\psi^-_{\xx,\o,\s'}\psi^-_{\yy,-\o,\s}\psi^+_{ \yy , 
-\o,\s'}\\
F_2&=\frac{1}{2}\sum_{\o,\s,\s'} \int 
\tilde 
w(\xx-\yy)\psi^+_{\xx,\o,\s}\psi^-_{\xx,\o,\s}\psi^-_{\yy,-\o,\s'}
\psi^+_{ \yy , 
-\o,\s'} \\
F_4&=\frac{1}{2}\sum_{\o,\s,\s'} \int 
\tilde 
w(\xx-\yy)\psi^+_{\xx,\o,\s}\psi^-_{\xx,\o,\s}\psi^-_{\yy,\o,\s'}
\psi^+_{ \yy , 
\o,\s'}\,.
\end{align*}
Here $\tilde 
w(\xx)$ is a short range interaction, with range $r_0$ and $\hat w(0)=1$. 
Setting 
\begin{equation*}
\tilde \j_{0,\xx}=\sum_\o \tilde\r_{\o,\xx}\,,\qquad \tilde \j_{1,\xx} 
=\sum_\o \o \tilde\r_{\o,\xx}, 
\end{equation*}
with $\tilde\r_{\o,\xx}=\sum_\s 
\psi^+_{\o,\s}\psi^-_{\o,\s}$, we get the WI for the fermionic correlations
\begin{align}
-ip_0 A_0\langle\hat{\tilde \j}_{0,\pp} &\hat\psi^-_{\kk+\pp,\sigma}
\hat\psi^+_{\kk,\sigma}\rangle_T + \nn\\
p v A_1&\langle\hat{\tilde \j}_{1,\pp} 
\hat\psi^-_{\kk+\pp,\sigma}
\hat\psi^+_{\kk,\sigma}\rangle_T=\label{wi1}\\
&\frac{1}{Z} \left[\langle\hat\psi^-_{\kk+\pp,\sigma}
\hat\psi^+_{\kk+\pp,\sigma}\rangle_T-\langle\hat\psi^-_{\kk,\sigma}
\hat\psi^+_{\kk,\sigma}\rangle_T\right] \nn
\end{align}
where $A_0=(1-\n_4-2\n_\r), A_1(1+\n_4-2\n_\r)$, $ \n_4= \bar g_4/4\pi v$ and 
$\n_\r=(\bar g_2-\bar g_1/2)/4\pi v$. Similarly, if $\tilde P_\o=-i 
p_0+\o v p$, the 
density correlations verify
\begin{align}
\tilde P_\o\langle\hat{\tilde\r}_{\pp,\o}& \hat{\tilde 
\r}_{-\pp,\o'}\rangle_T-\n_4 
\tilde P_{-\o}\langle\hat{\tilde \r}_{\pp,\o} \hat{\tilde 
\r}_{-\pp,\o'}\rangle_T-\label{aa}\\
&-2\n_\r \tilde P_{-\o}\langle\hat{\tilde \r}_{\pp,-\o} \hat 
{\tilde \r}_{-\pp,\o'}\rangle_T=-\d_{\o,\o'} 
\frac{\tilde P_{-\o}}{2 \pi Z^2} \nn
\end{align}
Note in the above WI the presence of the {\it anomalies}, that is te terms in 
$\n_\r$ and $\n_4$, which are linear in the couplings $\bar g_i$. The model 
differs from the Luttinger model for the presence of the $\bar g_1$ term; it is 
however defined so that it is invariant under the chiral phase transformation
\begin{equation*}
\psi^\pm_{\xx,\o,\s} \to e^{\pm i\a_{\xx,\o}}\psi^\pm_{\xx,\o,\s}
\end{equation*}
which imply, 
thanks to (\ref{aa}), that the density correlations can be explicitly computed 
even if the model is not solvable, see \cite{BFM,BFM2}.
We choose $\tilde w$ of the form $\tilde w(\xx)=\bar w(x^2+x_0^2/v^2)$ where  
$\bar w$ has range $r_0=2^{-h^*}$ 
and satisfies $\int d\xx |\tilde w(\xx)|=1$. It acts as an 
ultraviolet cut-off that allow us to integrate safely the scales 
$h\ge h^*$ and arrives to an effective potential $\overline{V}^{h^*}$, 
differing from $V^{h^*}$ discussed in the previous section by irrelevant 
terms. We can choose the 
bare parameters $\bar g_i, v$ of the reference model so that its running 
coupling constants differ from those of model (1) by exponentially 
decaying terms $O(2^{\th h})$ and the ratio of the $Z$ tends to $1$; this is 
achieved by choosing $ \bar g_{i} =g_{i,h^*}+O(\sqrt{\e}\tilde\l^2)) $ and $v= 
\sin p_F(1+O(\tilde\l\e^\th)+O(\tilde\l^2))$. This implies that
\begin{equation}
D(\pp)=\frac{Z_1^2}{Z^2}\langle\hat{\tilde \j}_{1,\pp}\hat{\tilde 
\j}_{1,-\pp}\rangle_T+R_0(\pp)
\end{equation}
where $Z_1$ is the current wave function normalization and $R_0(\pp)$ is a {\it 
continuous} function in $\pp$ (in contrast with the 
first addend in the r.h.s.); we use the WI $\lim_{p\to 0} \lim_{p_0\to 0} 
D(\pp)=0$ to fix $R_0(0)$ so that we get
\begin{equation*}
D(\pp)=\frac{Z_1^2}{ \pi Z^2 v  
v_1}\frac{[(1+\n_4+2\n_\r)+v_2^2(1-\n_4-2\n_\r)] 
p_0^2}{p_0^2+v_2^2 v^2 p^2}
\end{equation*}
with $v_2^2=\frac{(1+\n_4)^2-4\n_\r^2}{(1-\n_4)^2-4\n_\r^2}$, 
$v_1=(1+\n_4)^2-4\n_\r^2$. The identity 
\begin{equation*}
\langle j_{\pp} a^-_{\kk+\pp_F,\s}  
a^+_{\kk+\pp+\pp_F,\s} \rangle_T= Z_{1} \langle\tilde j_{1,\pp} 
\psi^-_{\kk+\pp,\s}\psi^+_{\kk,\s} \rangle_T
\end{equation*}
allows us to fix $Z_1,Z$; 
indeed comparing \eqref{wi1} with the WI for the 
chain
\begin{align}
-i p_0 \langle\hat\r_\pp \hat a^-_{\kk,\s} 
a^+_{\kk+\pp,\s}\rangle_{T}+&p\langle\hat j_\pp \hat
a^-_{\kk,\s} a^+_{\kk+\pp,\s}\rangle_{T}\\
=&\langle\hat a^-_{\kk,\s} a^+_{\kk,\s}\rangle_{T}-\langle\hat a^-_{\kk+\pp,\s} 
a^+_{\kk+\pp,\s}\rangle_{T}\nn
\end{align}
we get the consistency relations 
\begin{equation*}
\frac{Z_1}{Z}=v (1+\n_4-2\n_\r)\,.
\end{equation*}
Proceeding 
in a similar way for the susceptibility we obtain the expressions in the 
Theorem.

\section{Conclusions} We analyze non integrable generalizations of XXZ and 
the Hubbard chain in the low and high density regimes where the Luttinger 
description breaks down. Our methods are based on a multiscale decomposition of 
the propagator of the theory and are able to take into account, in a rigorous 
and quantitative way, the irrelevant terms normally neglected in RG analysis. 
Our main conclusion is that no qualitative difference between solvable and non 
solvable models are seen in exponents and conductivity at zero temperature, even 
in regions where Luttinger liquid description is not valid and the physics is 
completely dominated by irrelevant terms. In particular the anomalous critical 
exponents vanishes or not depending on the spinless or spinful nature of 
fermions, and the Drude weight tends to the same non interacting values in both 
cases. 

It is common belief that the Drude weight is zero in the presence of non 
integrable interactions, while still non zero for integrable ones, as soon as 
$T\not=0$. However, the fact 
that at $T=0$ integrability breaking  terms do not produce any difference in the transport 
properties, even in regimes where irrelevant terms dominate, 
makes it also possible a scenario where
breaking of integrability effects in transport may matters only at not too low 
temperature \cite{Le}.

\section{Appendix A: Flow of the running coupling constants in the quadratic 
regime}
We give some extra detail on the flow of the r.c.c. in the quadratic regime. 
Note that at $r=0$ and $T=0$ we have
\begin{itemize}
\item empty band case: $p_F=0$, $e(k)=-\cos k+1$, and
\begin{equation*}
g(\xx)=\chi(x_0>0)\int_ {-\pi}^\pi \frac{dk}{2\pi} e^{-i 
kx-e(k)x_0}
\end{equation*}
\item filled band case: $p_F=\pi$, $e(k)=-\cos k-1$, and
\begin{equation*}
g(\xx)=-\chi(x_0\le 0)\int_ {-\pi}^\pi \frac{dk}{2\pi} e^{-i 
kx-e(k)x_0}
\end{equation*}

\end{itemize}
Therefore all the graphs with order greater than $1$ with two external lines are 
vanishing if computed at the Fermi points and $\e=0$. Indeed all one particle 
reducible graphs are vanishing due to the support properties of the propagator. 
This implies that there is always a closed fermionic loop which vanishes as the 
propagator is proportional to $\chi(x_0>0)$ or $\chi(x_0\le 0)$. At 
first order there are two contributions: the tadpole graph at $\e=0$ contributes 
only to $\n$ and  gives $2 \l \hat w(0) p_F/\pi$ with $p_F=0,\pi$; the other 
graph is vanishing for non local interactions (local potential does not 
contribute)  since $v(\xx-\yy) g(\xx-\yy)$ is proportional to 
$v(x,y)\d_{x,y}=0$.

The flow equations for $i_h,\d_h$ have the form $i_{h-1}=i_h+\b_i^h$,  
$\d_{h-1}=\d_h+\b_i^h$. In the spinless case the fact that there are no quartic  
running coupling constants produce an improvement of $O(2^{h \th})$ with respect 
to the dimensional bound . As we noticed above all the contributions with two 
external lines computed at the Fermi points are vanishing for $\e=0$, except 
the tadople which contributes only to $\n_h$. There is therefore a gain $\e 
2^{-h}$ in the beta function for $z,\d$, and a further gain $2^{h\th}$  (due to 
the irrelevance of the quartic terms if the order is greater then $1$ and to 
the fact that the derivative can be applied on the interaction at first order), 
so we get $|i _h|,|\d_h|\leq \sum_{k=h}^1 C|\l|\e 2^{-k} 2^{k\th}$ and finally 
$z_{h^*}, \d_{h^*}=O(\l \e^\th)$. The same argument can be used for the 
renormalization of the chemical potential $\n_h$ and $\n_0$ is the tadpole plus 
$\sum_{h=h^*}^1 \l  2^h \e 2^{-h} 2^{\th h}=O(\l \e)$; as a consequence the 
shift of the critical chemical potential is linear in $\l$ as stated in the 
Theorem.

In the spinful case, the contributions at first order to the flow of $i_h,\d_h$ 
give $\tilde\l\sum_{h\ge h^*} \e 2^{-h} 2^{\th h}\le C\e^{\th}\tilde\l$ for the 
same reason as in the spinless case. There is however no gain due to the 
irrelevance of the interaction at larger orders so that they  give $\tilde\l^2 C 
\sum_{h\ge h^*} \e 2^{-h}\le C\tilde\l^2$ as the quartic terms are now relevant. 
Finally, the value of $\n$ is the tadpole plus $\sum_{h=h^*}^1 \tilde\l  2^h \e 
2^{-h}=O(\tilde\l \sqrt{\e})$.

\section{Appendix B: Flow of the running coupling constants in the linear 
regime}

In the spinless case the beta functions for $\l_h$ and $v_h$ are convergent and 
asymptotically vanishing, $|\b_\l^h|\le C \frac{\l^2_{h}}{v_{h^*} } 
\frac{2^h}{v^2_{h^*}}$, $|\b_\d^h|\le C \frac{\l_{h}^2}{v_{h^*}^2 
}\frac{2^h}{v^2_{h^*}}$. Assuming inductively that $|\l_h|\le C \l 
\e^{1/2+\th}$ 
and using that $\frac{2^h}{v^2_{h^*}}\le 2^{h-h^*}$ one gets so that 
\be 
|\l_{h-1}-\l_{h^*}|\le \sum_{k=h}^{h^*} \e^{1+2\th} \frac{\l^2}{v_{h^*}} 
2^{k-h^*}\le  C \l^2 \e^{1/2+\th} 
\ee 
and $v_{-\io}=v_{h^*}+O(\frac{\l_{h^*}^2}{v_{h^*}^2})\sim \e^{\frac12}$. 
Moreover $\frac{Z_{h-1}}{Z_h}=1+\b^1_z+\b^2_z$ where $\b^2$ contains the 
contributions from the irrelevant terms, like the quadratic corrections to the 
dispersion relation, and is $O(\l\frac{\g^h}{v_{h^*}} )$. Finally at first order 
$\d_h$ has contibutions only from non-local terms, the derivative is applied on 
the interaction and is bounded by $\l/v \sum_{k\le h^*} 2^k$ either in spinful 
and spinless case.


\begin{thebibliography}{999999}
\bibitem{1}
T. Kinoshita, T. Wenger, and D. S. Weiss, Nature 440, 900 (2006).
\bibitem{2}
S. Hofferberth, I. Lesanovsky, B. Fischer, T. Schumm, and
J. Schmiedmayer, Nature 449, 324 (2007).
\bibitem{2xx}
A. Mazurenko, Christie S. Chiu, Geoffrey Ji, Maxwell F. Parsons, Márton 
Kanász-Nagy, Richard Schmidt, Fabian Grusdt, Eugene Demler, Daniel Greif  Markus 
Greiner Nature 545, 462--466 (2017)
\bibitem{ko} 
Essler, F. H. L.; Frahm, H., Goehmann, F., Kluemper, A., Korepin, V. E., The 
One-Dimensional Hubbard Model. Cambridge University Press (2005).
\bibitem{5}
X. Zotos, F. Naef, and P. Prelovsek, Phys. Rev. B ˇ 55, 11029 (1997).
\bibitem{6}
T. Prosen and E. Ilievski, Phys. Rev. Lett. 111, 057203 (2013).
\bibitem{6aa}
J.-S. Caux, R. Hagemans, J.-M. Maillet J., Stat. Mech. P09003 (2005) 
\bibitem{gla} 
A. Imambekov, T. L. Schmidt,  L. I. Glazman Rev. Mod. Phys 84, 1253 (2012)
\bibitem{gla1}
R. G. Pereira, J. Sirker, J.-S. Caux, R. Hagemans, J. M. Maillet, 
S. R. White, I. Affleck J., Stat. Mech. P08022 (2007) 
\bibitem{1a}
M.A. Cazalilla,	Phys. Rev. Lett., vol. 97, 15, 156403 (2003)
\bibitem{2a}
J. Lancaster, A. Mitra, Phys. Rev. E, 81, 6, 061134 (2010
\bibitem{3a}
T. Sabetta, G.  Misguich, Phys. Rev. B,  88, 24, 245114 (2013)
\bibitem{4a}
D. Bernard,  B.Doyon, Journal of Statistical Mechanics 6, 6, 064005 (2016)
\bibitem{5a}	
B. Bertini, M. Collura, J. De Nardis, M. Fagotti,
Phys. Rev. Lett. 117, 20, 207201 (2017)
\bibitem{6a}
E. Langmann, J. L. Lebowitz, V. Mastropietro, P. Moosavi, 
Commun. Math. Phys. 349, 551 (2017); Phys. Rev. B 95, 235142 (2017)
\bibitem{7a}
C. Karrasch, T. Prosen, F. Heidrich-Meisner,
Phys. Rev. B 95, 060406 (2017)
\bibitem{8a}
E. Ilievski, J. De Nardis, Phys. Rev. Lett. 119, 2, 02060 (2017)
\bibitem{9a}
C. Karrasch, New J. Phys. 19, 033027 (2017)
\bibitem{Ha} 
F.D.M. Haldane, Phys.Rev.Lett. 45, 1358--1362 (1980);  J. Phys. C. 14, 
2575--2609 (1981).
\bibitem{M1} 
D. C.. Mattis, V.Mastropietro, The Luttinger model (World Scientific 2014)
\bibitem{BFM} 
G. Benfatto,  P. Falco, V. Mastropietro, Phys. Rev. Lett. 104, 
075701 (2010); Comm. Math. Phys. 330, 1, 153-215 (2014); Comm. Math. Phys 330, 
1,217-282 (2014)
\bibitem{BoL} 
F. Bonetto, J.L. Lebowitz, L. Rey-Bellet, Mathematical Physics 
2000, Edited by A. Fokas, A. Grigoryan, T. Kibble and B. Zegarlinsky, Imprial 
College Press, 128-151 (2000) 
\bibitem{8}
J. V. Alvarez and C. Gros, Phys. Rev. Lett. 88, 077203
(2002); Phys. Rev. B 66, 094403 (2002).
\bibitem{10}
P. Jung and A. Rosch, Phys. Rev. B 76, 245108 (2007).
\bibitem{12}
Heidrich-Meisner, F.; Honecker, A.; Brenig, W.,
Eur. Phys. J. Special Topics 151, 135-145 (2007)
\bibitem{13}
D. Heidarian and S. Sorella, Phys. Rev. B 75, 241104 (2007)
\bibitem{13a}
J. Sirker, R. G. Pereira, and I. Affleck, Phys. Rev. Lett.
103, 216602 (2009); Phys. Rev. B 83, 035115 (2011)
\bibitem{15}
R. Steinigeweg, J. Herbrych, X. Zotos, W. Brenig, Phys. Rev. Lett. 116, 
017202 (2016)
\bibitem{x1}A. Giuliani, Vi Mastropietro, M. Porta
Phys. Rev. B 83, 195401 (2011)
\bibitem{x2}D.L. Boyda,,  V.V Braguta, M.I.  Katsnelson, M-V. Ulybyshev,
Phys. Rev. B 94, 085421 (2016)
\bibitem{BFM2} 
G.Benfatto P. Falco V. Mastropietro Comm. Math.Phys 330(1), 217-282 (2014)
\bibitem{MP} 
V. Mastropietro, M. Porta J. Stat., Phys. (2017)
\bibitem{S1} 
J. Sirker, Int. J. Mod. Phys. B, 26, 1244009 (2012)
\bibitem{PZ} 
C. Psaroudaki, X. Zotos, J. Stat. Mech. (2016) 063103
\bibitem{S} 
H.J. Schulz, arxiv9302006
\bibitem{BoM} 
F. Bonetto, V. Mastropietro, Annales Henri Poincar\'e 17 (2), 459-495 (2016)
\bibitem{Le} J. Lebowitz,  J. Scaramazza arxiv 1801.07153 
\end{thebibliography}
\end{document}